\documentclass[pra,twocolumn,showpacs,amsmath,amssymb,superscriptaddress]{revtex4}
\usepackage{graphicx}
\usepackage{dcolumn}
\usepackage{bm}
\usepackage{appendix}

\begin{document}
\title{Resolving all-order method convergence problems for atomic physics applications}
\author{H. Gharibnejad}
\affiliation {Department of Physics, University of Nevada, Reno, Nevada 89557, USA.}
\author{E. Eliav}
\affiliation {Department of Chemistry, Tel Aviv University, Tel Aviv, Israel.}
\author{M. S. Safronova}
\affiliation {Department of Physics, University of Delaware, Newark, Delaware 19716, USA.}
\author{A. Derevianko}
\affiliation {Department of Physics, University of Nevada, Reno, Nevada 89557, USA.}

\begin{abstract}
The development of the relativistic all-order method where all single, double, and partial triple
excitations of the Dirac-Hartree-Fock wave function
are included to all orders of perturbation theory led to many important results for study of fundamental symmetries,
 development of atomic clocks,
ultracold atom physics, and others, as well as provided recommended values of many atomic
 properties critically evaluated for their accuracy for
large number of monovalent systems.  This approach requires iterative solutions of the
 linearized coupled-cluster equations leading to
convergence issues in some cases where correlation corrections are particularly large or lead
 to an oscillating pattern. Moreover, these issues also lead to
similar problems in the CI+all-order method for many-particle systems. In this work, we
 have resolved most of the known convergence problems by
applying two different convergence stabilizer methods, reduced linear equation (RLE) and direct
inversion of iterative subspace (DIIS). Examples are
presented for B, Al, Zn$^+$, and Yb$^+$. Solving these convergence problems greatly expands the number of atomic species that can be treated with the all-order methods  and is anticipated to facilitate
many interesting future applications.
\end{abstract}

\maketitle

\section{Introduction}
\label{Sec:Introduction} The coupled-cluster (CC) method has been successfully applied to solve quantum many-body problems in quantum chemistry
~\cite{Èiz66, Bart89} as well as computational
 atomic~\cite{LinMor86} and nuclear physics~\cite{Coe60}. A relativistic linearized variant
 of the coupled-cluster method (which is numerically symmetric and is generally referred to as ``all-order method'') was developed for atomic physics
 applications in Refs.~\cite{BluJohLiu89a,BluJohLiu89b,BluJohSap91}. It is one of the most accurate methods currently
 being used in the atomic structure
calculations. This approach was extremely
 successful and led to accurate predictions of energies, transition amplitudes,
hyperfine constants, polarizabilities, $C_3$ and $C_6$ coefficients, isotope shifts, and other properties of monovalent atoms, as well as the
calculation of parity-nonconserving (PNC) amplitudes in Cs, Fr, and Ra$^+$ (see
\cite{BluSapJon92,SafJohDer99,DerJohSaf99,SafJoh00fr,SafJoh07,PalJiaSaf09} and references therein). Further development of the all-order approach, that
included triple excitations and non-linear terms yielded  the most precise evaluation of the PNC amplitude in Cs \cite{PorBelDer09,PorBelDer10} and consequent
re-analysis of Cs experiment \cite{WooBenCho97}. This work  provided the most accurate low-energy test of the electroweak sector of  the Standard
Model to date, placed constraints on a variety of new physics scenarios beyond the SM, and, when  combined with the results of
high-energy collider experiments, confirmed the energy dependence (or ``running'') of the
electroweak force over an energy range spanning four orders of magnitude (from $\sim$10~MeV to $\sim$100~GeV). All-order method was also used for development of ultra-precise atomic clocks
\cite{FlaDzuDer08,BelSafDer06,SafJiaSaf10,JiaAroSaf09,SafJiaAro10}, ultracold atom and quantum information studies
\cite{Der10,MorDzuDer11,RavDerBer06,AroSafCla07,SafWilCla03}  and many other applications. We refer the reader to review \cite{SafJoh07} for details
of the all-order method and its applications. The all-order method is also used as a part of the CI+all-order approach for study of more complicated
systems~\cite{SafKozJoh09}.

 The all-order method requires iterative solutions of the
 linearized coupled-cluster equations leading to
convergence issues in some cases when correlation corrections are very large or produce
  an oscillating iterative pattern. The initial guess of the solution
 is based on the low-order perturbation theory. Therefore, if
 high-order correlation corrections are large, initial guess is very
 poor leading to very slow convergence or failure of the straightforward iterative scheme.
  In addition, initial non-linear CC equations may have more than one solution, so a convergence to non-physical
  solutions may occur. Several such problems have been identified over
  the years and led to failure to apply all-order approach for
 many important applications. For example, all or almost all of the low-lying  $nd$ and $nf$
 states of B, Al, Zn$^+$, Cd$^+$, Hg$^+$, and Yb$^+$ do not converge if
standard Jacobi-type iterative procedure is applied.
 In the case of Yb$^+$, even core equations do not converge.
 Convergence problems also cause complete failure of the all-order
 approach for super-heavy elements, such as element 113 (eka-Tl).
All these convergence issues in monovalent systems lead to the same problems in the application of the CI+all-order approach \cite{SafKozJoh09} to
the corresponding divalent systems, such as Al$^+$, Hg, Yb, etc. since this method required prior solution of LCCSD equations for one-particle
orbitals. There are several interesting present applications of these atoms and ions that require high-precision calculations possible
 with all-order techniques. For example, several
of these systems are used or proposed for optical clocks \cite{ChoHumKoe10,FlaDzuDer08,RosHumSch08,DzuDer10,TamWeyLip09} requiring precise knowledge
of the blackbody radiation (BBR) shift which is hard to accurately measure. BBR shift is a leading source of uncertainties for many of the atomic
clock schemes. Yb is used for an ongoing PNC experiment \cite{TsiDouFam09} as well as studies of degenerate quantum gases
\cite{IvaKhrHan11,TasNemBau10} owing to a number of available isotopes. The best available Yb PNC amplitude value is only accurate to 20\%
\cite{PorRakKoz95}.

The convergence issues that arise in the solutions of eigenvalue equations have a long history in general quantum chemistry and several methods have
been developed to address them~\cite{PurBar81,Trucks88,Pul80,Pul82,MosElKa01,Harrison03,EliavXIH05}. Most of these methods are based on the
fundamental idea of effective reduction of original large functional space and solution of the projected to the reduced (Krylov) subspace of the
simplified equations. This idea was implemented for the first time in a quantum chemical application by Lanczos \cite{Lan50}, who facilitated a
partial diagonalization of a large matrix by transforming to a much smaller Krylov subspace, followed by a matrix triangularization procedure. In the
present work, we consider two such convergence techniques, namely, reduced linear equation (RLE) \cite{PurBar81,Trucks88} and the direct inversion of
iterative subspace (DIIS) \cite{Pul80,Pul82}. Both of the methods use approximate solutions obtained from few last iterations as Krylov reduced
functional subspace onto which the linearized equations are projected and in which the projected system of equations is solved. The convergence of
the methods is based on the construction of error vectors. Different choices of the error vectors lead to different implementation of the methods.
Among the most popular error vectors are: 1) the difference of subsequent iterations and 2) ``true'' error vector (e.g. difference between exact
solution and it's approximation). In our work, both the convergence methods use the same best least squares approximation to the true error vector
and thus are rather relative. Moreover, our variant of DIIS can be regarded as a ``symmetric'' version of RLE (see below). However, while DIIS method
is chosen to minimize the error vector in the least-squares sense, the RLE differs from it by requiring that this vector within the basis vanishes.
We formulate here implementations of the RLE\ and DIIS methods\ for our variant of the coupled-cluster equations and test these stabilizer methods on
several specific examples, in which we were able to resolve the convergence problems listed above.
We also studied the effectiveness of these two techniques in solving specific types of the convergence problems as well as accelerating convergence
in all other cases. Acceleration of convergence is particulary important for further CI+all-order use since it
 requires solving all-order equations
for a large number of one-particle orbitals.

  Below, we briefly outline the essence of the convergence
stabilization procedures. In the coupled-cluster method, the desired exact wave function $|\psi\rangle$ is obtained by applying
 (a yet unknown operator) $\exp(T)$ on some reference
wave function $|\phi\rangle$, for example, the Dirac-Hartree-Fock (DHF) wavefunction. For a closed-shell system with $N$ electrons,
the cluster operator $T=\sum T_p$ (where $p=1,2,3...,N$) has the form:%
\begin{align}\label{Eq:clusteramp}
T_p=1/p!\sum_{mn..ab..} \rho_{mn...ab...}a_m^\dag a_n^\dag...a_aa_b...
\end{align}
Here, orbitals $m, n...$ are single-particle excited states; $a, b ...$ are core states which are occupied in $|\phi\rangle$; $\rho$'s are cluster
amplitudes (also called excitation coefficients) and $a^\dag$ and $a$ are creation and annihilation operators with respect to the quasi-vacuum
state~$|\phi\rangle$. Finally, $p$ is the number of core electrons excited when applying $T_p$ to $|\phi\rangle$. In the linearized coupled-cluster
single-double (LCCSD) method,  only $T_1$ and $T_2$ are retained and non-linear terms in the expansion of $\exp(T)$ are truncated.
 The  LCCSD equations are
conventionally solved by  an iterative scheme, symbolically written as $\rho^{(n+1)}=F(\rho^{(n)})$, $F$ being specified later in
Section~\ref{Sec:LCCSD}. In this paper, this type of straightforward iteration procedure is referred to as the conventional iterations scheme (CIS).

 Both RLE and DIIS convergence stabilization procedures
form $\rho^{(n+1)}$ solution as the linear combination of cluster amplitudes  ($\rho^{(n)}, \rho^{(n-1)},..., \rho^{(n-l)}$)
 accumulated from  $l$ previous CIS iterations.
  Further details of the LCCSD method and RLE and DIIS schemes are discussed in Sections \ref{Sec:LCCSD} and \ref{Sec:Methods of Convergence}.\\
\begin{figure}
\begin{center}
\includegraphics[scale=.35]{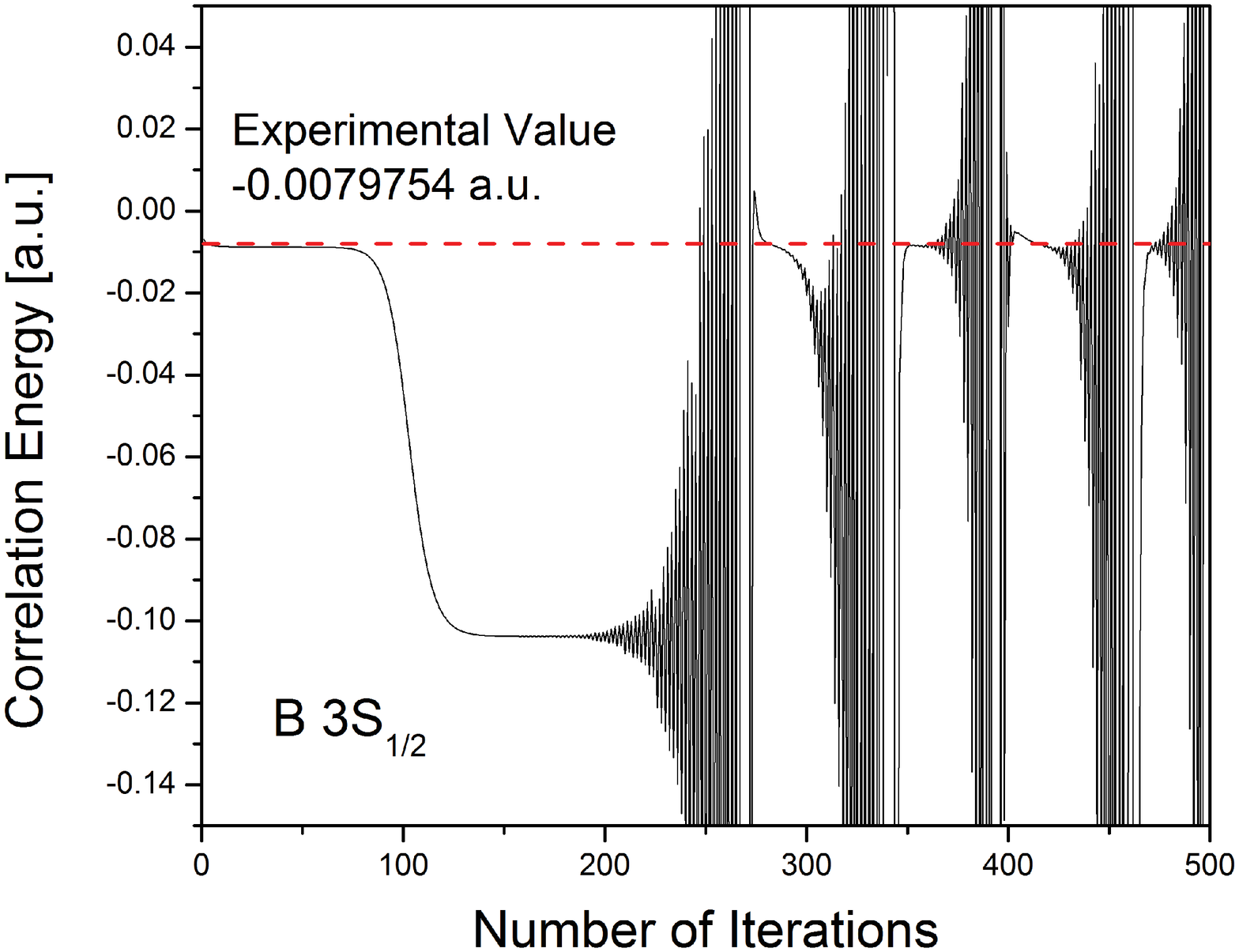}
\end{center}
\caption[]
  {(Color online) The failure of the LCCSD straightforward iteration procedure for the $3s$ state
  in boron. The calculated correlation energy is plotted as a function of the iteration number.
 The dashed (red) line indicates the value of the experimental correlation energy.}\label{Fig:B3sno}
\end{figure}

An example of failed conventional iteration procedure is shown in Fig.~\ref{Fig:B3sno}, where we plot the LCCSD correlation energy,
 $\delta E$, as a function of a
number of valence LCCSD iterations for the $3s$   state of boron. The experimental correlation energy ($-0.0079754$~a.u.) is indicated by the
horizontal dashed line. It is  computed by subtracting DHF energy from the experimental result. The LCCSD $3s$ correlation energy diverges from the
experimental values dramatically and begins to oscillate after a number of iterations. The convergence criteria is set to terminate the iteration
procedure when the relative difference between two
 consecutive iterations is reduced below 0.00001. The convergence is not reached even after 500 iterations. As demonstrated below, this
problem is completely resolved by the use
  of either RLE or DIIS procedures and convergence to the above criteria is reached within 30 iterations.


This paper is organized as follows: in Section~\ref{Sec:LCCSD}, we describe the LCCSD method and the conventional  iteration procedure (CIS) of
solving the LCCSD equations.
 In Section~\ref{Sec:Methods of Convergence}, we formulate RLE and DIIS schemes for  LCCSD equations. In Section~\ref{Sec:Results},
  we analyze performance of the RLE and DIIS
 schemes for various cases. Finally, in Section~V, we draw the conclusions.
\section{Linearized SD Coupled-Cluster  Method}
\label{Sec:LCCSD} In the present implementation of the CC method, the exact valence wave function $|\psi_v\rangle$ is obtained from the lowest-order
DHF state,
\begin{align}\label{Eq:lowestDHF}
 |\phi_v\rangle=a^\dagger_v|0_c\rangle\, ,
\end{align}
 by applying a wave operator $\Omega=N[\exp(T)]$~\cite{LinMor86}:
\begin{align}\label{Eq:exactwave}
|\psi_v\rangle=\Omega|\phi_v\rangle\,,
\end{align}
where $|0_c\rangle$ is the core DHF state and $N[...]$ designates the normal product of operators with respect to a closed-shell core.
 Taking into account only the $T_1$ and $T_2$ terms in Eq.~(\ref{Eq:clusteramp}), and truncating $\Omega$ past the linear terms in the
 expansion of the exponential leads to the LCCSD ansatz for the wave operator
\begin{align}\label{Eq:LSDwaveoperator}
\Omega \simeq &1+\sum_{ma}\rho_{ma}a_m^\dag a_a +\frac{1}{2}\sum_{mnab}\rho_{mnab}a_m^\dag a_n^\dag a_b a_a + \notag \\ \notag & \ \sum_{m\neq
v}\rho_{mv}a_m^\dag a_v+\sum_{mna}\rho_{mnva}a_m^\dag a_n^\dag a_a a_v\\ =&1+S_c+D_c+S_v+D_v.
\end{align}
Here, $S_c$ and $D_c$ ($S_v$, $D_v$) are the core (valence) single and double terms, respectively.

 To find the cluster amplitudes (or excitation coefficients) $\rho$, we need to specify the Hamiltonian. In our approach, we use the
 Hamiltonian \cite{BluJohSap91} $H=H_0+G$:
\begin{align}\label{Eq:Hamiltonian}
H=\sum_i \varepsilon_i N[ a_i^\dag a_i] + \frac{1}{2}\sum_{ijkl}g_{ijkl} N[a_i^\dag a^\dag_ja_la_k] \, ,
\end{align}
where $H_0$ is the one-electron lowest-order DHF Hamiltonian and $G$ is the residual Coulomb interaction. Indices $i$, $j$, $k$, and $l$ range over
all possible single-particle orbitals, and $g_{ijkl}$ are the two-body Coulomb matrix elements. A set of coupled equations for the cluster operators
$(T)_n$:
$$
 (T_c)_1 = S_c,~ (T_v)_1 = S_v,~ (T_c)_2 = D_c,~\textrm{and}~  (T_v)_2 = D_v$$
 may be found from the Bloch equation~\cite{LinMor86}. For
monovalent systems~\cite{DerEmm02}:
\begin{eqnarray}\label{Eq:Blochcore}
(\varepsilon_v-H_0)(T_c)_n&=&\{QG\Omega\}_{\textrm{connected, n}}, \\ \label{Eq:Blochval} (\varepsilon_v+\delta
E_v-H_0)(T_v)_n&=&\{QG\Omega\}_{\textrm{connected, n}},
\end{eqnarray}

\noindent where $\delta E_v=\langle \phi_v|G\Omega|\phi_v \rangle$ is the valence correlation energy and $Q=1-|\phi_v \rangle\langle \phi_v|$
 is the projection operator. Note that Eq.~(\ref{Eq:Blochcore}) contains only the core cluster operators, while Eq.~(\ref{Eq:Blochval}) contains both core and
 valence cluster operators. The core equations (\ref{Eq:Blochcore}) are solved first, and the resulting CC core
 amplitudes are subsequently frozen and used in the valence equations (\ref{Eq:Blochval}).

 The summations over the magnetic quantum numbers $m$ in Eqs.~(\ref{Eq:Blochcore}) and ~(\ref{Eq:Blochval}) are performed analytically.
After the angular reduction, the equation for the reduced single core cluster amplitudes $\rho(ma)$ takes form~\cite{SafJohDer99,Saf00}:
\begin{eqnarray}
&&(\varepsilon_a-\varepsilon_m)\rho(ma)= \label{Eq:reducedsing} \\
&&\delta_{\kappa_m\kappa_a}\{\sum_{nb}\delta_{\kappa_n\kappa_b}\sqrt{\frac{[j_b]}{[j_a]}}Z_0(mban)\rho(nb) \nonumber \\
&&-\sum_k\sum_{ncb}\frac{(-1)^{j_a+j_b+j_c+j_n}}{[j_a][k]} Z_k(cbna)\rho_k(nmcb)\nonumber \\
&&+\sum_k\sum_{rnb}\frac{(-1)^{j_a+j_b+j_r+j_n}}{[j_a][k]} Z_k(mbrn)\rho_k(rnab)\}\,. \nonumber
\end{eqnarray}
Here, $[j]=2j+1$, $\kappa$ is the relativistic angular momentum quantum number,  $\rho(ma)$ and $\rho_k(mnab)$ are reduced single and double cluster
amplitudes, $X_k(mnab)$ are  reduced two-body Coulomb matrix elements, and
\begin{eqnarray*}
Z_k(mnab)&=&X_k(mnab)\\&+&\sum_{k'}[k]\left(\begin{array}{ccc}
                      j_m & j_a & k \\
                      j_n & j_b & k' \\
                    \end{array}
                  \right)X_{k'}(mnba).
\end{eqnarray*}

The equations for the reduced double core cluster amplitudes $\rho_k(mnab)$ are given by:
\begin{eqnarray} \label{Eq:reduceddob}
&& \hspace{-0.5cm}\left( \varepsilon _{ab}-\varepsilon _{mn} \right) \rho_k(mnab) =X_k(mnab) \rule[-2ex]{0em}{0ex}\\
 &&\hspace{-0.5cm}+\sum_{cd}\sum_{l,k^{\prime }}A_1 X_l(cdab)\rho _{k^{\prime }}(mncd)  \nonumber \\
&&\hspace{-0.5cm}+\sum_{rs}\sum_{l,k^{\prime }}A_2 X_l(mnrs)\rho _{k^{\prime }}(rsab)\nonumber \\
&&\hspace{-0.5cm}+\left[ \sum_rX_k(mnrb)\rho (ra)\delta _{\kappa _r\kappa _a}+\sum_cX_k(cnab)\rho (mc)\delta _{\kappa _m\kappa_c} \right. \nonumber \\
&&\hspace{-0.5cm}-\left.\sum_{rc} \frac{(-1)^{j_c+j_r+k}}{\left[ k\right] }Z_k(cnrb){\widetilde{\rho} }%
_k(mrac)\right]  + \left[
\begin{array}{c} a \leftrightarrow b \\
                 m \leftrightarrow n \end{array}
 \right], \nonumber
\end{eqnarray}
where $A_i$  are angular coefficients given in ~\cite{Saf00} and $\varepsilon_{ij}=\varepsilon_i+\varepsilon_j$. The valence equations have exactly the same form as the core equations with the replacement of index $a$ by the valence
index $v$ everywhere and an addition of the valence correlation energy $\delta E_v$ into the energy difference on the left-hand side, i.e.,
$(\varepsilon_a-\varepsilon_m)\rho(ma) \longrightarrow (\varepsilon_v-\varepsilon_m+\delta E_v)\rho(mv)$.

 Implementation of
the RLE and DIIS procedures requires rewriting the equations for the cluster amplitudes in a specific vector form.
 We introduce the vector notation:
\begin{align}\
\notag
 \textbf{t}=&\left(\begin{array}{c}
                      \rho(ma) \\
                      \rho_k(mnab) \\
                    \end{array}
                  \right)\, ,
\end{align}
where $\rho(ma)$ and $\rho_k(mnab)$ are to be understood as columns composed of all  amplitudes for single and double excitations, respectively, i.e.,
for all possible values of $m$, $n$, $a$, $b$, and $k$ indexes. Then, the core equations given by Eqs.~(\ref{Eq:reducedsing}) and (\ref{Eq:reduceddob}) may be
combined as
\begin{align}\label{Eq:combinedcore}
\textbf{D}\cdot \textbf{t}=-\textbf{a}-\Delta \cdot \textbf{t}\,,
\end{align}
where
\begin{align}\notag
\textbf{a}=&-\left(\begin{array}{c}
                      0 \\
                      X_k(mnab) \\
                    \end{array}
                  \right),~~~
                   \textbf{D}=\left(\begin{array}{c}
                      \varepsilon_a-\varepsilon_m \\
                      \varepsilon_{ab}-\varepsilon_{mn} \\
                    \end{array}
                  \right),
\end{align}
and $\Delta \cdot \textbf{t}$ includes all terms on the right-hand sides of the Eqs.~(\ref{Eq:reducedsing}) and (\ref{Eq:reduceddob})
 except for $X_k(mnab)$, which is included in $\textbf{a}$.

Valence equations may be written in the same way with
\begin{align}\
\notag
 \textbf{t}=&\left(\begin{array}{c}
                      \rho(mv) \\
                      \rho_k(mnvb) \\
                    \end{array}
                  \right)\,
\end{align}
and
\begin{align}\notag
\textbf{a}=&-\left(\begin{array}{c}
                      0 \\
                      X_k(mnvb) \\
                    \end{array}
                  \right), ~~~
                   \textbf{D}=\left(\begin{array}{c}
                      \varepsilon_v-\varepsilon_m + \delta E_v\\
                      \varepsilon_{vb}-\varepsilon_{mn} + \delta E_v\\
                    \end{array}
                  \right).
\end{align}
The main difference between the core and valence equations for the implementation of the RLE and DIIS is the dependence of the valence  array $D$ on the
iteration number, since $\delta E_v$ is recalculated after every iteration. In the core case, $D$ remains constant.

 Solving Eq.~(\ref{Eq:combinedcore}) for $\textbf{t}$ gives
\begin{align}\label{Eq:main}
\textbf{t}=-\textbf{D}^{-1}(\textbf{a}+\Delta \cdot \textbf{t})\,.
\end{align}
The above equation can be solved iteratively as
\begin{align}\label{Eq:ite}
\textbf{t}^{(m+1)}=-\textbf{D}^{-1}(\textbf{a}+\Delta \cdot \textbf{t}^{(m)})\,.
\end{align}
The iteration usually starts by inserting $\textbf{t}^{(0)}=0$ on the right hand side of Eq.~(\ref{Eq:ite}) and finding $\textbf{t}^{(1)}$.
 As we demonstrated in Fig.~\ref{Fig:B3sno}, convergence of this straightforward iterative scheme is occasionally very slow or fails altogether.
  The convergence methods that we develop in the next section will alleviate such problems and  lead to faster convergence rates.
\section{RLE and DIIS}
\label{Sec:Methods of Convergence}

In this section, we formulate  implementation of RLE and DIIS methods for the
  LCCSD equations (\ref{Eq:main}) discussed in the previous section.
Both methods are two-step procedures. In the first step, a few iterative solutions $\mathbf{t}^{(i)}$ of Eq.~(\ref{Eq:ite}) are found (same as the CIS). In the
second step, a linear combination of these $\mathbf{t}^{(i)}$ is used to find the next best  solution of Eq.~(\ref{Eq:ite}). The new answer is then used for
another  initialization of the CIS and the two steps are repeated until convergence to specified criteria is reached. In this section, we present the
general RLE and DIIS formulas and derive their explicit form for the LCCSD equations.

 After accumulating   $m+1$ iteratively found solutions, $\textbf{t}^{(1)}, \textbf{t}^{(2)}, ..., \textbf{t}^{(m+1)}$,
next best approximation can be found as their linear combination,
\begin{align}\label{Eq:linearcomb}
\textbf{t}^{[m+1]}=\sum _{i=1}^{m} \sigma_i \textbf{t}^{(i)}=\sigma \cdot \textbf{T}\,.
\end{align}
The quantities  $\sigma_i$ are the weights  that have to be determined  by solving a system of  equations constructed from previously found $m+1$
CIS solutions. We note that $\textbf{t}^{(m+1)}$ is not included in the linear combination (\ref{Eq:linearcomb}), but is
   used to find $\sigma_i$ coefficients. Therefore, we use the notation $\textbf{t}^{[m+1]}$ instead
   of $\textbf{t}^{(m+1)}$ to distinguish between the $m+1^{th}$ solution found through the use of RLE/DIIS methods and the initial
CIS result, respectively.

 Both direct inversion of iterative space (DIIS) and reduced linear equation
(RLE) methods seek to minimize the error between the iteratively found solutions of Eq.~(\ref{Eq:main}) and the exact answer.
 The error minimization is the basis for finding the appropriate $\sigma_i$ to form the approximate solution $\textbf{t}^{[m+1]}$.
 Both methods also use a least square approach to the error minimization.
 Since the exact answer is unknown, approximations are used  in the minimization process.
 The approximate solution, as mentioned before, is constructed as a linear combination of a series of iteratively found solutions.
 The difference between the DIIS and  the RLE methods is in the assumptions they make in order to minimize the errors.
 Further details of the
difference between the two methods and derivations of the DIIS/RLE formulas can be found in the Appendix A.

We rewrite Eq.~(\ref{Eq:combinedcore}) as
\begin{align}\label{Eq:linear}
\textbf{a}+(\Delta+\textbf{D})\textbf{t}=\textbf{a}+\textbf{B}\textbf{t}=0.
\end{align}
The DIIS formula for determining $\sigma_i$ is given by  Eq.~(\ref{Eq:diisA}):
\begin{align}\label{Eq:diis}
\textbf{T}^T\textbf{B}^T\textbf{a}+\textbf{T}^T\textbf{B}^T\textbf{B}\textbf{T}\sigma=0\,.
\end{align}
The  RLE formula for determining $\sigma_i$ is given by Eq.~(\ref{Eq:rleA}):
 \begin{align}\label{Eq:rle}
\textbf{T}^T(\textbf{a}+\textbf{BT}\sigma)=0\,.
\end{align}
Both Eqs.~(\ref{Eq:diis}) and (\ref{Eq:rle}) can be written as a system of $m$ equations:
\begin{align}\label{Eq:alphaR}
\boldsymbol{\alpha}+\textbf{R}\sigma=0\,.
\end{align}
 Solving the above system of equations for $\sigma$ can be easily done with standard linear algebra methods.
 The resulting coefficients $\sigma_i$ are substituted into Eq.~(\ref{Eq:linearcomb})  to obtain best new approximate solution $\textbf{t}^{[m+1]}$.

 Next, we write $\mathbf{R}$ and $\boldsymbol{\alpha}$ of Eq.~(\ref{Eq:alphaR}) in their explicit forms for DIIS and RLE methods.
 Substituting $\Delta-\textbf{D}$ for $\textbf{B}$  into DIIS equation  (\ref{Eq:diis}) yields for the $\sigma_i$
\begin{align}\nonumber
\mathbf{t}^{^{T}\left(  i\right)  }(\Delta+\textbf{D})^Ta+\mathbf{t}^{^{T}\left(  i\right)  }(\Delta+\textbf{D})^T(\Delta+\textbf{D})\mathbf{t}^{\left(  i\right) }\sigma_i=0\,.
\end{align}
 Using Eq.~(\ref{Eq:ite}), we find that $\Delta \cdot \textbf{t}^{(i)}=-(\textbf{D} \cdot \textbf{t}^{(i+1)}+\textbf{a})$.
 Replacing the dot products involving $\Delta$ with ones involving $\textbf{D}$ yields
 explicit form of DIIS matrix for core orbitals
\begin{align}
&R_{ij}    =\sum_{k}D_{kk}a_{k}\left( t_{k}^{\left(  i+1\right)  }+t_{k}^{\left( j+1\right)  }-t_{k}^{\left(  i\right)  }-t_{k}^{\left(  j\right)
}\right)
+\sum_{k}\left(  a_{k}\right)  ^{2} \nonumber\\
&+\sum_{k}D_{kk}^{2}\left(  t_{k}^{\left(  i\right)  }t_{k}^{\left(
j\right)  }+t_{k}^{\left(  i+1\right)  }t_{k}^{\left(  j+1\right)  }%
-t_{k}^{\left(  i+1\right)  }t_{k}^{\left(  j\right)  }-t_{k}^{\left( i\right)  }t_{k}^{\left(  j+1\right)  }\right)  \nonumber \\
&\alpha_{i}    \mathbf{=}\sum_{k}a_{k}D_{kk}\left(  t_{k}^{\left(  i\right) }-t_{k}^{\left(  i+1\right) }\right)  -\sum_{k}\left(  a_{k}\right)
^{2}\,.\label{Eq:assigndiis}
\end{align}

The RLE equations for core orbitals are obtained by repeating the same steps as for the DIIS approach but starting from Eq.(\ref{Eq:rle}). The
resulting RLE equations for $\textbf{R}$ and $\boldsymbol{\alpha}$ are
\begin{align}\label{Eq:assignrle}
R_{ij}=&\sum _{kl} t^{(i)}_k (\Delta_{kl}+D_{kl})t^{(j)}_l \nonumber \\
=&\sum _{k} [t^{(i)}_k D_{kk}t^{(j)}_k-t^{(i)}_k D_{kk} t^{(j+1)}_k-a_k t^{(j)}_k ]\,, \nonumber \\ \alpha_i=&\sum _k t_k^{(i)} a_k\,.
\end{align}

We noted in the previous  section that $\textbf{D}$ depends on the correlation energy, $\delta E_v$, in the case of the valence equations leading to
the dependence of $\textbf{D}$ on the iteration number.  Therefore,  the substitution $\textbf{D}\rightarrow \textbf{D}^{(i)}$ must
  be made to rewrite the DIIS and RLE equations above for the valence orbitals.
  To derive the final form of the equations, we have to introduce a somewhat arbitrary dot product and normalization definitions.
   The explicit form of the core RLE equations is obtained by substituting the expressions  for  $\textbf{D}$, $\textbf{a}$, and $\textbf{t}$ from the previous section
   into Eq.~(\ref{Eq:assignrle}):
\begin{align}\label{Eq:assignrleexplicitcore}
 R_{ij}=&
\sum_{ma}\left(  \varepsilon_{a}-\varepsilon_{m}\right)  \rho^{\left( i\right)  }\left(  ma\right)  [\rho^{\left(  j\right)  }\left(
ma\right)-\rho^{\left(  j+1\right)  }\left(  ma\right)]  \nonumber \\  & + \sum_{L}\sum_{mnab}\frac{1}{[L]}\left(  \varepsilon_{ab}-\varepsilon
_{mn}\right) \rho_{L}^{\left(  i\right)  }\left(  mnab\right) \nonumber \\
&\times  [\rho _{L}^{\left(  j\right)  }\left(  mnab\right)-\rho _{L}^{\left( j+1\right) }\left( mnab\right)]-\alpha_{i}\, , \nonumber\\
 \alpha_{i}
 =&-\sum_{L}\sum_{mnab}\frac{1}{[L]}X_{L}\left(  mnab\right) \rho_{L}^{\left( i\right) }\left( mnab\right)\, .
\end{align}
 RLE equations for valence case are given by
\begin{align}\label{Eq:assignrleexplicit}
 R_{ij}=&
\sum_{ma}\left(  \varepsilon_{a}-\varepsilon_{m}+\delta E_v ^{(j)}\right)  \rho^{\left( i\right)  }\left(  ma\right) \nonumber \\
&\times [\rho^{\left( j\right) }\left(  ma\right)-\rho^{\left(  j+1\right)  }\left(  ma\right)]  \nonumber \\ & + \sum_{L}\sum_{mnab}\frac{1}{[L]}
\frac{1}{[j_v]} \left( \varepsilon_{ab}-\varepsilon _{mn}+\delta E_v ^{(j)}\right)  \rho_{L}^{\left(  i\right)  }\left(  mnab\right) \nonumber \\
&\times [\rho _{L}^{\left(  j\right) }\left( mnab\right)-\rho _{L}^{\left(  j+1\right)  }\left(  mnab\right)]-\alpha_{i}\, ,\nonumber \\  \alpha_{i}
=&-\sum_{L}\sum_{mnab}\frac{1}{[L]}X_{L}\left( mnab\right) \rho_{L}^{\left(  i\right)  }\left(  mnab\right)\, ,
\end{align}
 where $[L]=2L+1$.

 The implementation of the RLE and DIIS methods proceeds as follows. In step one, our code makes a limited number, $m+1$, of LCCSD iterations using the CIS.
  This is done to find $m+1$ series of single and double cluster amplitudes, $\rho^{(i)}(ma)$ and $\rho_L^{(i)}(mnab)$ that are then saved.
 In step two, a separate subroutine applies the DIIS or RLE equations to these stored cluster amplitudes to find the appropriate $\textbf{R}$ and $\boldsymbol{\alpha}$ matrices.
 The $m$-dimensional linear equation (\ref{Eq:alphaR}) is solved for $\sigma$.
  The next best solution of the LCCSD equations is then found by substituting $\sigma$ into Eq.~(\ref{Eq:linearcomb}).
  These two steps are repeated until convergence  is reached according to a specified criteria.
In the next section, we discuss the results of the application of the DIIS and RLE procedures to the solution of the LCCSD equations in the cases
that do not converge or converge to non-physical answers with the conventional iteration scheme.
\section{Results and discussion}
\label{Sec:Results}

\begin{table}
\caption{\label{tab1} Convergence tests of the LCCSD equations with CIS, RLE, and DIIS methods for B and Al. CIS is the conventional iterations
scheme (no convergence stabilizer). RLE5 designates RLE convergence method with 5 pre-stored iterations. Last column gives resulting correlation
energy in a.u. *Cases where maximum number of iterations allowed during run was reached.}
\begin{ruledtabular}\begin{tabular}{llcccc}
\multicolumn{1}{c}{Atom } & \multicolumn{1}{c}{State} & \multicolumn{1}{c}{Method} &\multicolumn{1}{c}{\# of iter.} & \multicolumn{1}{c}{Converged?}
&
\multicolumn{1}{c}{$\delta E_v (a.u.)$}\\
\hline
B   &   Core    &   CIS &   21  &   Yes &     \\
    &       &   RLE5    &   9   &   Yes &     \\
    &       &   DIIS5   &   15  &   Yes &     \\  [0.3pc]
B   &   $2p_{1/2}$   &   CIS &   18  &   Yes &   -0.0293907  \\
    &       &   RLE5    &   13  &   Yes &   -0.0293906  \\
    &       &   DIIS5   &   13  &   Yes &   -0.0293908  \\  [0.3pc]
B   &   $3s$  &   CIS &   70* &   No  &   -0.0091643  \\
    &       &   RLE4    &   70* &   No  &   -0.0070438  \\
    &       &   DIIS4   &   70* &   No  &   -0.0089454  \\
    &       &   RLE5    &   30  &   Yes &   -0.0089491  \\
    &       &   DIIS5   &   22  &   Yes &   -0.0089472  \\
    &       &   DIIS7   &   31  &   Yes &   -0.0089488  \\  [0.3pc]
B   &   $3p_{1/2}$   &   CIS &   44  &   Yes &   -0.0056292  \\
    &       &   RLE5    &   23  &   Yes &   -0.0056294  \\
    &       &   DIIS5   &   19  &   Yes &   -0.0056284  \\
    &       &   RLE8    &   25  &   Yes &   -0.0056295  \\
    &       &   DIIS8   &   24  &   Yes &   -0.0056293  \\[0.3pc]
B   &   $3d_{3/2}$   &   CIS &   85* &   No  &   -0.0884489  \\
    &       &   DIIS7   &   71  &   Yes &   -0.0007535  \\
    &       &   RLE8    &   66  &   Yes &   -0.0007536  \\
    &       &   DIIS8   &   51  &   Yes &   -0.0007533  \\[0.3pc]
Al  &   Core    &   CIS &   13  &   Yes &     \\
    &       &   RLE5    &   8   &   Yes &     \\[0.3pc]
Al  &   $3p_{1/2}$   &   CIS &   16  &   Yes &   -0.0245810  \\
    &       &   RLE5    &   11  &   Yes &   -0.0245798  \\
    &       &   DIIS5   &   14  &   Yes &   -0.0245811  \\[0.3pc]
Al  &   $4s$  &   CIS &   20  &   Yes &   -0.0079907  \\
    &       &   RLE5    &   13  &   Yes &   -0.0079906  \\
    &       &   DIIS5   &   18  &   Yes &   -0.0079905  \\[0.3pc]
Al  &   $3d_{3/2}$   &   CIS &   70* &   No  &   -0.0209573  \\
    &       &   DIIS6   &   89  &   Yes &   -0.0208637  \\
    &       &   RLE8    &   86  &   Yes &   -0.0208662  \\
    &       &   DIIS8   &   49  &   Yes &   -0.0208662  \\[0.3pc]
Al  &   $4d_{3/2}$   &   CIS &   70* &   No  &   -0.0213727  \\
    &       &   RLE8    &   300*    &   No  &   0.0022280   \\
    &       &   DIIS8   &   81  &   Yes &   0.0022478   \\
    &       &   DIIS9   &   91  &   Yes &   0.0022477   \\[0.3pc]
Al  &   $4d_{5/2}$   &   CIS &   70* &   No  &   -1.3775005  \\
    &       &   RLE8    &   165 &   Yes &   0.0022737   \\
    &       &   DIIS8   &   97  &   Yes &   0.0022710   \\
    &       &   DIIS9   &   73  &   Yes &   0.0022712
 \end{tabular}
\end{ruledtabular}
\end{table}

\begin{figure}
\begin{center}
\includegraphics[scale=0.59]{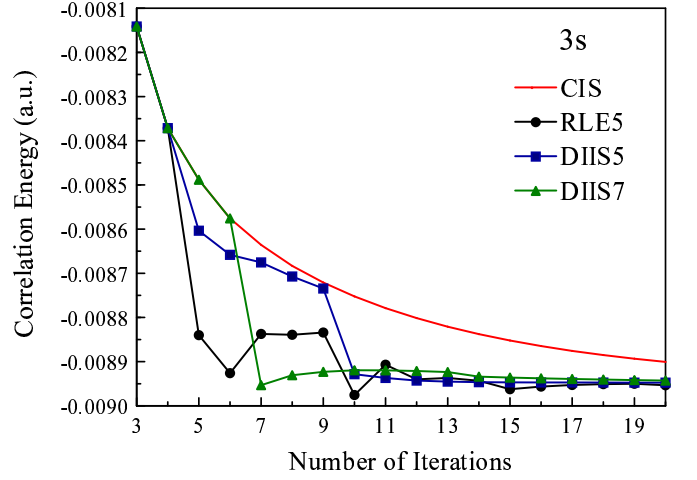}
\end{center}
\caption[]
  {(Color online) Comparison of the RLE5, DIIS5, and DIIS7 schemes for the $3s$ state of boron.  The correlation energy is given in a.u. }\label{fig1}
\end{figure}

In this section, we  study and compare the convergence characteristics of the DIIS and the RLE methods.
We include a number of test cases in four
different systems, B, Al, Zn$^+$, and Yb$^+$ which have a large number of states that do not converge with the conventional iterative scheme (CIS).
We also test the ability of the RLE and the DIIS to accelerate convergence in the cases where the CIS does converge. The main purpose of these tests is to
provide general guidelines of how to accelerate or to achieve convergence using the RLE and the DIIS methods. The conclusions and observations of this section may
be extrapolated to other systems for both all-order and CI+all-order approaches.

The summary of B and Al convergence tests is given in Table~\ref{tab1}. We
find that convergence patterns for two fine-structure multiplet states, for example $3p_{1/2}$ and $3p_{3/2}$ states, are generally very similar.
Therefore, we
 list only $np_{1/2}$ and $nd_{3/2}$ states with the exception of the $4d$ states of Al. Tests were performed for both states of the multiplet as an
 additional check, since similar results are expected.
 The results are given for the $2p_{1/2}$, $3s$, $3p_{1/2}$, and $3d_{3/2}$ states of B and the $3p_{1/2}$, $4s$, $3d_{3/2}$, $4d_{3/2}$ and $4d_{5/2}$
states of Al.  The resulting LCCSD correlation energy is listed in the last column of the table in a.u. The convergence method is specified in the
third column. CIS refers to the initial straightforward iteration scheme. RLE5 designates the RLE convergence method with 5 pre-stored CIS
iterations. Similarly, DIIS8 refers to the DIIS convergence scheme with 8 pre-stored iterations. The fourth column indicates the iteration number at
the end of the run. Cases where the maximum number of iterations allowed during the run was reached are marked with asterisk. In these cases,
convergence did not occur. The convergence criteria was set to terminate the iteration procedure when the relative difference between two
 consecutive iterations is reduced below 0.00001. The same convergence criteria is used in all valence test runs.
 Only the core and the $2p$ states of B converge with the CIS. In the case of Al, all $nd$ states do not converge with the CIS. All of the cases in Table~\ref{tab1}
 converge with the DIIS8. We note that we did not list the RLE5 and the DIIS5 results for many of the $nd$ states
because convergence was not achieved. In cases where all methods lead to convergence, both the RLE and the DIIS have accelerated convergence rates relative to the CIS.
\begin{figure} [ht]
\begin{center}
\includegraphics[scale=0.59]{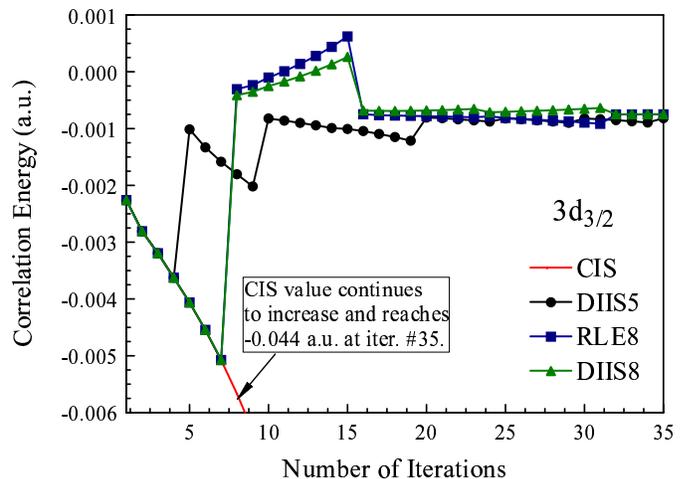}
\end{center}
\caption[]
  {(Color online) Comparison of the DIIS5, RLE8, and DIIS8 schemes for the $3d_{3/2}$ state of boron.  The correlation energy is given in a.u. }\label{fig2}
\end{figure}

We may draw two general conclusions from our tests:
\begin{enumerate}
\item{ If a particular LCCSD run converges with the CIS, then the RLE5 appears to be the most efficient method in accelerating the convergence.}
\item{ If a particular LCCSD run does {\em not} converge with the CIS, the DIIS8 or the DIIS9 appear to be the most efficient in attaining and accelerating the convergence.}
\end{enumerate}

We note that these two rules are not absolute, but they serve to be good initial guidelines. Our further tests on other (much heavier) systems confirmed these
guidelines.  We note that the RLE5 is not sufficient to achieve convergence for most of the divergent cases. The only exception in Table~\ref{tab1} is
the $3s$ state of B. However, while the CIS never converges to our standard criteria for the $3s$ state, it nearly converges to correct result before
exhibiting diverging and oscillating pattern of Fig.~\ref{Fig:B3sno}. In this case, accumulation of only 5 iterations is sufficient. However, in the case of
the $nd$ states, the CIS is never close to converging to a correct number and subsequently the RLE5 does not work. Occasionally, DIIS9 may achieve convergence where
DIIS8 would not. Using even larger number of stored iterations does not improve convergence or efficiency. DIIS10-DIIS12 runs for the $4d$
 states converged to non-physical answers in two instances, but to correct results in all other cases. Number of
 iterations varied significantly from case to case.
The results of all converged runs listed in Table~\ref{tab1} are consistent within the convergence criteria, as expected.

We implemented the DIIS/RLE strategies for two separately developed LCCSD codes. The calculations were carried out using two different finite basis sets, the B-splines of Ref.~\cite{JohBluSap88} and the dual-kinetic-basis sets of Ref.~\cite{BelDer08}. Even though the basis sets and the convergence criteria used for each code made slight differences in the values, the general observations on the convergence patterns remains the same.

\begin{table}
\caption{\label{tab2} Convergence tests of the LCCSD equations with CIS, RLE, and DIIS methods for Zn$^+$ and Yb$^+$. CIS is the conventional
iterative scheme (no convergence stabilizer). DIIS8 designates the DIIS convergence method with 8 pre-stored iterations. Last column gives resulting
correlation energy in a.u. *Cases where maximum number of iterations allowed during run was reached.}
\begin{ruledtabular}\begin{tabular}{llcccc}
\multicolumn{1}{c}{Atom } & \multicolumn{1}{c}{State} & \multicolumn{1}{c}{Method} &\multicolumn{1}{c}{\# of iter.} & \multicolumn{1}{c}{Converged?}
&
\multicolumn{1}{c}{$\delta E_v (a.u.)$}\\
\hline
Zn$^+$ &   $5p_{1/2}$   &   CIS &   39  &   Yes &   -0.0066119  \\
    &       &   DIIS9   &   12  &   Yes &   -0.0066088  \\  [0.5pc]
Zn$^+$ &   $4d_{3/2}$   &   CIS &   70* &   No  &   -0.0977215  \\
    &       &   RLE5    &   67  &   Yes &   -0.0045508  \\
    &       &   DIIS8   &   18  &   Yes &   -0.0045511  \\  [0.5pc]
Zn$^+$ &   $4d_{5/2}$   &   CIS &   70* &   No  &   -0.1149058  \\
    &       &   RLE5    &   93  &   Yes &   -0.0045266  \\
    &       &   DIIS8   &   18  &   Yes &   -0.0045267  \\  [0.5pc]
Zn$^+$ &   $5d_{3/2}$   &   CIS &   70* &   No  &   -0.1936330  \\
    &       &   RLE5    &   28  &   Yes &   -0.0018929  \\
    &       &   DIIS8   &   18  &   Yes &   -0.0018942  \\  [0.5pc]
Zn$^+$ &   $4f_{5/2}$   &   CIS &   200* &   No  &  -0.0008697  \\
    &       &   DIIS8   &   200*    &   No  &   -0.0007949  \\
    &       &   DIIS9   &   153 &   Yes &   -0.0007948  \\  [0.5pc]
Zn$^+$ &   $4f_{7/2}$   &   CIS &   70* &   No  &   -0.0007970  \\
    &       &   DIIS8   &   173 &   Yes &   -0.0007891  \\
    &       &   DIIS9   &   195 &   Yes &   -0.0007891  \\  [0.5pc]
Yb$^+$ & Core      &   CIS &       &   No  &       \\
    &       &   RLE5    &   12  &   Yes &     \\
    &       &   DIIS5   &   12  &   Yes &
\end{tabular}
\end{ruledtabular}
\end{table}

\begin{figure} [ht]
\begin{center}
\includegraphics[scale=0.59]{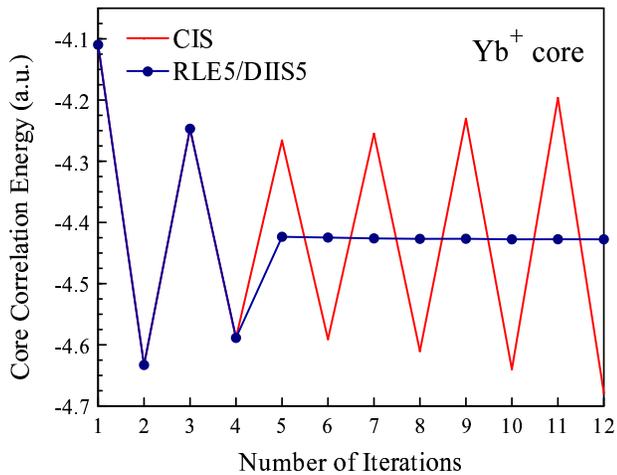}
\end{center}
\caption[]
  {(Color online) Comparison of the CIS, RLE5, and DIIS5 schemes for the Yb$^+$ core.  The correlation energy is given in a.u.
  RLE5 and DIIS5 data appear identical at this scale and are shown as a single curve. }\label{fig4}
\end{figure}
We illustrate different convergence patterns of the RLE and the DIIS methods for the $3s$ and $3d_{3/2}$ states of boron in Figs.~\ref{fig1} and \ref{fig2}.
In Fig.~\ref{fig1}, the values of the correlation energies for the
$3s$ states of boron are obtained from different schemes that are listed on the graph. RLE5, DIIS5, and DIIS7 results after $N=20$ interactions are indistinguishable at this plot scale and are not shown. These
schemes converge after 30, 22, and 31 iterations, respectively. While the CIS results appear close to converged value,
convergence was never reached and correlation energy began to oscillate as illustrated in Fig.~\ref{Fig:B3sno}. The RLE5 and the DIIS5 results are identical
to the CIS ones for the first four iterations. The 5th value is different for three of the schemes as this $(m+1)$-th value (see Eq.~(\ref{Eq:linearcomb})) is
replaced by the RLE or the DIIS predictions for RLE5 and DIIS5. We observe that these predictions are significantly closer to converged
result than the 5th CIS iteration. After that, the RLE5 and DIIS5 results are sharply adjusted at $N=10$ when the second call to the RLE/DIIS stabilizer codes is made. The DIIS7 behavior is similar to the one just described, except that it accumulates 7 iterations before the DIIS procedure is invoked and now the 7th value gets much closer to the final answer.

\begin{table*}
\caption{\label{tab3} Comparison of B, Al, Zn$^+$, and Yb$^+$ removal energies (in cm$^{-1}$) with experiment \cite{SanMarYou06}. Rows labeled
``Dif.'' give relative difference with experimental values in \%. }
\begin{ruledtabular}
\begin{tabular}{lcccccccc}
\textbf{B}   &   $2p_{1/2}$   &   $2p_{3/2}$   &   $3s$  &   $3p_{1/2}$   &   $3p_{3/2}$   &   $3d_{3/2}$   &   $3d_{5/2}$   &       \\
\hline
Expt.   &   -66928  &   -66913  &   -26888  &   -18316  &   -18314  &   -12160  &   -12160  &       \\
SD  &   -67049  &   -67035  &   -27105  &   -18497  &   -18495  &   -12494  &   -12494  &       \\
Dif.  &   -0.18\%   &   -0.18\%   &   -0.81\%   &   -0.99\%   &   -0.99\%   &   -2.7\%   &   -2.7\%   &       \\  [0.5pc] \hline
 \textbf{Al}     & $3p_{1/2}$
&   $4s$  &   $3d_{3/2}$   &   $3d_{5/2}$   &   $4d_{3/2}$   &   $4d_{5/2}$   &       &       \\
\hline
Expt.   &   -48278  &   -22931  &   -15843  &   -15842  &   -6045   &   -6041   &       &       \\
SD  &   -48271  &   -23069  &   -17295  &   -17289  &   -6652   &   -6647   &       &       \\
Dif.  &   0.02\%    &   -0.60\%   &   -8.4\%    &   -8.4\%    &   -9.1\%    &   -9.1\%    &       &       \\  [0.5pc] \hline
\textbf{Zn$^+$}   &   $4s$  &   $4p_{1/2}$   &   $4p_{3/2}$   &   $5p_{1/2}$   &   $4d_{3/2}$   &   $4d_{5/2}$   &   $5d_{3/2}$   &   $4f_{7/2}$   \\
\hline
 Expt.   &   -144691 &   -96027  &   -95157  &   -43360  &   -47950  &   -47902  &   -26913  &   -27606  \\
SD &   -144684 &   -96221  &   -95352  &   -43421  &   -47929  &   -47880  &   -26898  &   -27633  \\
SDpT  &   -145232 &   -96559  &   -95679  &   -43492  &   -47994  &   -47946  &   -26929  &   -27633  \\
Dif. (SD) &   0.14\%    &   0.20\%    &   0.19\%    &   0.24\%    &   0.11\%    &   0.11\%    &   0.09\%    &   -0.02\%   \\
Dif. (SDpT)  &   -0.23\%   &   -0.15\%   &   -0.15\%   &   0.08\%    &   -0.02\%   &   -0.03\%   &   -0.02\%   &   -0.01\%   \\  [0.5pc] \hline
\textbf{Yb$^+$} &   $6s$  &   $6p_{1/2}$   &   $6p_{3/2}$   &   $7s$  &   $5d_{3/2}$   &   $5d_{5/2}$  &   $5f_{5/2}$   &   $5f_{7/2}$   \\
\hline
Expt.   &   -98207  &   -71145  &   -67815  &   -43903  &   -75246  &   -73874  &   -27704  &   -27627  \\
SD &   -98961  &   -71016  &   -67480  &   -44060  &   -76141  &   -74700  &   -28080  &   -28062  \\
SDpT  &   -99107  &   -71084  &   -67592  &   -44115  &   -77764  &   -76317  &       &       \\
Dif. (SD) &   -0.77\%   &   0.18\%    &   0.49\%    &   -0.36\%   &   -1.19\%   &   -1.12\%   &   -1.36\%   &   -1.57\%   \\
Dif. (SDpT)  &   -0.91\%   &   0.09\%    &   0.33\%    &   -0.48\%   &   -3.2\%   &   -3.2\%   &       &       \\
\end{tabular}
\end{ruledtabular}
\end{table*}

In Fig.~\ref{fig2}, the values of the correlation energies obtained from CIS, DIIS5, RLE8, and DIIS8 are plotted for the
$3d_{3/2}$ states of boron. The RLE8 and the DIIS8 results after $N=35$ appear identical on the graph at this scale and are not shown. The RLE8 and the DIIS8
converge to our criteria after 66 and 51 iterations, respectively. Very similar behavior of the RLE8 and the DIIS8 is observed, with the RLE8 energy
oscillations being slightly larger after the RLE subroutine pass. However, other tests show that the RLE8 in general converges slower, sometimes
dramatically so, than the DIIS8. The CIS values diverge completely and increase rapidly. The DIIS5 seems to be converging at $N=35$, but
does not in fact reach selected criteria even after 100 iterations.

The summary of the selected Zn$^+$ and Yb$^+$ convergence tests is presented in Table~\ref{tab2}. The results are given for the $5p_{1/2}$, $4d_{3/2}$,
$4d_{5/2}$, and $5d_{3/2}$, $4f_{5/2}$, and $4f_{7/2}$ states of Zn$^+$ and the Yb$^+$ core. The $4s$ and $4p_{j}$  states of Zn$^+$ and  the $6s$,
$6p_j$, $7s$, and $5d_j$ states of Yb$^+$ converge with the CIS, so we have omitted these results from the table. However, it is worth pointing out that RLE5 accelerates convergence for
all these states compared to the CIS. Table~\ref{tab2} demonstrates that the DIIS reduces number of iterations for the $5p_{1/2}$ states by a factor of 3 or better. Zn$^+$ and Yb$^+$
tests confirm our conclusions (1) and (2), on the previous page. We were unable to achieve convergence for higher $7p_j$ states in Yb$^+$. This problem is not
present in Zn$^+$, where LCCSD for the $5p$ states converges even with CIS as shown in Table~\ref{tab2}. Perhaps other convergence approaches are needed to
resolve this issue.

The case of Yb$^+$ core is particularly interesting, since core iterations generally converge well with the CIS. Yb$^+$ core is
an exception, however, most likely due to very large $4f$ shell contributions that lead to oscillation of the correlation energy. We plot the CIS, the RLE5,
and the DIIS5 results for the Yb$^+$ core correlation energy in Fig.~\ref{fig4}. The RLE5 and the DIIS5 results appear identical at this scale and are shown as
 a single curve. Both methods are successful at fixing the CIS's oscillation problem.

The comparison of the B, Al, Zn$^+$, and Yb$^+$ removal energies with experiment ~\cite{SanMarYou06} is given in Table~\ref{tab3}. Rows labeled
``Dif.'' give relative difference with experimental values in \%. The energies here are given in cm$^{-1}$. Most of these states did not converge
with the CIS, so it is important to establish the accuracy of this approach for such cases. Breit interactions and contributions from higher partial
waves are also included. The B and Al ionization potentials, B $2p_{3/2}$, and Al $4s$ SD energies are in agreement with experiment. We consider only
monovalent states for all of these systems. The SD approximation does not account for mixing with the core-excited states such as $3s3p^2$ in Al.
Therefore, larger disagreement with experiment is expected in cases where mixing with these hole-two-particle states is large. A particular example
is the $3d$ and $4d$ states of Al. The lower $3s^2nd$ levels heavily mix with the $3s3p^2\, ~^2D$ levels. However, the mixing coefficient for this
configuration never exceeds 30\%. As a result, these levels are distributed over several lower $nd$ levels, resulting in two sets of levels being
listed as $3s^2 4d~ ^2\!D$~\cite{KauMar91,SanMarYou06} ($[y\,^2\!D]$ and $[^2D]$). In Table~\ref{tab3}, we compare the $4d$ results with the  second
sets of levels ($[^2D]$).

We also included partial valence triples perturbatively (LCCSDpT) to investigate if the LCCSDpT method would improve the theory-experiment agreement for Zn$^+$ and Yb$^+$. This
method is described in detail in \cite{SafJohDer99}. Since triple equations are not explicitly iterated in this approach, implementation of the RLE
and the DIIS method is exactly the same as in the SD code. Convergence tests of the LCCSDpT method exhibit essentially the same pattern as the tests of the LCCSD
method discussed above, and a similar number of iterations was generally required for LCCSD and LCCSDpT calculations for the same states run with the same
parameters.

As shown in Table~\ref{tab3}, we find an excellent agreement of all Zn$^+$ data with experiment. Inclusion of perturbative triples somewhat improves the agreement with
 experiment for  most states. The accuracy decreases for Yb$^+$, as expected, owing to much softer and heavier core and strong mixing of
 monovalent states with one-hole-two-particle states in this system. Nevertheless, for Yb$^+$ the average accuracy for removal energies is at the level of 1\% (see Table~\ref{tab3}.)

 \section{Conclusion}
\label{conclusion} We have successfully implemented the RLE and DIIS  convergence techniques  in the LCCSD and LCCSDpT methods for high-precision
atomic many-body calculations. Most of the convergence problems  were resolved using these methods. Acceleration of convergence was demonstrated for
all cases where all-order equations converge with straightforward iteration scheme.  Numerous tests were performed to establish general
recommendations for the RLE/DIIS use for various purposes.  We find that if particular case converges with CIS, RLE5 appears to be the most efficient
in achieving and accelerating convergence.
 If particular case does not converge with CIS, DIIS8 or DIIS9 appear to be the most efficient in accelerating convergence.
 Solving these convergence problems greatly expands the number of atomic species that can be treated with the all-order methods  and is anticipated to facilitate
many interesting future applications for studies of fundamental symmetries as well as  atomic clock and ultracold atom research.
\appendix
\section{}
 Derivations of general formulas in this Appendix  mainly follows  Appendix of Ref.~\cite{PurBar81}.
 Consider solving a general linear equation of the form
\begin{align}\label{Eq:linearA}
\textbf{a}+\textbf{B}\textbf{t}=0\,,
\end{align}
which is a system of linear equations of dimension $k$ with vector $\textbf{t}$ being the exact solution that we would like to find.
 We make the best approximation to the exact solution by using $m(<k)$
  nonorthogonal and linearly independent vectors $\textbf{T}=(\textbf{t}^{(1)}, \textbf{t}^{(2)}, ..., \textbf{t}^{(m)})$,
  where each $\textbf{t}^{(i)}$ is a $k$-dimensional vector. We find this best approximation as a linear combination of $\textbf{t}^{(i)}$'s:
\begin{align}\label{Eq:linearcombA}
\textbf{t}^{[m+1]}=\sum _{i=1}^{m} \sigma_i \textbf{t}^{(i)}=\sigma \cdot \textbf{T}\,.
\end{align}
Here, $\sigma_i$ are the weights of the optimized solution that needs to be determined. We note that $\textbf{t}^{(m+1)}$ is not included in the
linear combination (\ref{Eq:linearcombA}), but is
   used to construct matrices such as shown in Eqs.(\ref{Eq:assignrleexplicitcore}) and (\ref{Eq:assignrleexplicit}). Therefore, $\textbf{t}^{(m+1)}$ needs to be also found through the CIS. Therefore, we use the notation $\textbf{t}^{[m+1]}$ instead
   of $\textbf{t}^{(m+1)}$ to distinguish between the $m+1^{th}$ solution found through the use of RLE/DIIS methods and the
CIS result, respectively.

First, we try to derive an ideal equation to find $\sigma_i$'s as if we know the exact solution to Eq.~(\ref{Eq:linearA}). To find the best
approximation, we need to minimize the error between the approximate and the exact answers. To this end, we use the least square optimization
approach.
 The error is $\textbf{e}=\textbf{t}-\textbf{t}^{[m+1]}$.
 The least square optimization of $E=\textbf{e}^T\textbf{e}$ with respect to $\sigma$ then yields
\begin{align}\label{}
\frac{\partial E}{\partial \sigma}=-2\textbf{T}^T(\textbf{t}-\textbf{T}\cdot\sigma)=0\,.
\end{align}
After solving for $\sigma$ and substituting it in Eq.(\ref{Eq:linearcombA}), we get
\begin{align}\label{Eq:mainA}
\textbf{t}^{[m+1]}=\textbf{T}(\textbf{T}^T\textbf{T})^{-1}\textbf{T}^T\textbf{t}\,.
\end{align}
\indent However, not knowing what the exact solution $\textbf{t}$ is, the above formula is of little use.
 The DIIS and RLE are based on replacing $\textbf{t}$ with approximations.
Substituting $\textbf{t}^{[m+1]}$ instead of $\textbf{t}$ in Eq.(\ref{Eq:linearA}), will make Eq.(\ref{Eq:linearA}) inhomogeneous:
\begin{align}\label{Eq:errorA}
\textbf{a}+\textbf{B}\textbf{t}^{[m+1]}=\textbf{a}+\textbf{B}\textbf{T}\cdot\sigma=\epsilon\,.
\end{align}
where $\epsilon$ is a vector with constant elements.\\
\indent The difference between the RLE and DIIS methods is in their choice of error to minimize, $\textbf{e}$. The DIIS takes the error to be
$\epsilon$ of Eq.~(\ref{Eq:errorA}). Then to get the best approximation, we need to minimize $E=\epsilon^T\epsilon$ with respect to $\sigma$:
\begin{align}\label{}
\frac{\partial E}{\partial \sigma}=2(-\textbf{B}\textbf{T})^T(\textbf{a}+\textbf{B}\textbf{T}\sigma)=0\,.
\end{align}
Therefore, the coefficients $\sigma$ that lead to the best approximation satisfy the DIIS equation:
\begin{align}\label{Eq:diisA}
\textbf{T}^T\textbf{B}^T\textbf{a}+\textbf{T}^T\textbf{B}^T\textbf{B}\textbf{T}\sigma=0\,.
\end{align}
\\
\indent
The RLE requires that the best least square approximation $\epsilon^{[m+1]}$ to $\epsilon$ vanishes in the space of $T$. Following the structure of Eq.~(\ref{Eq:mainA}):
 \begin{align}\label{}
\epsilon^{[m+1]}=\textbf{T}(\textbf{T}^T\textbf{T})^{-1}\textbf{T}^T\epsilon=\textbf{T}(\textbf{T}^T\textbf{T})^{-1}\textbf{T}^T(\textbf{a}+\textbf{BT}\sigma)=0\,.
\end{align}
Since $\textbf{T}$ is made of linearly independent vectors, $\epsilon^{[m+1]}$ is only zero if:
\begin{align}\label{Eq:rleA}
\textbf{T}^T(\textbf{a}+\textbf{BT}\sigma)=0\,.
\end{align}
\\ \indent
Eqs.~(\ref{Eq:diisA}) and (\ref{Eq:rleA}) for DIIS and RLE, respectively, correspond to Eqs.~(\ref{Eq:diis}) and (\ref{Eq:rle}) in the paper.

\section*{Acknowledgments}
The work of H.G. and A.D. was supported in part by the US National Science Foundation Grant  No.\ PHY-9-69580. The work of M.S.S. was supported in part by National Science
Foundation Grant  No.\ PHY-07-58088.


\end{document}